\documentstyle[aps,twocolumn,prl,epsf]{revtex}

\begin{document}

\newcommand{\be}{\begin{equation}}
\newcommand{\ee}{\end{equation}}

\draft

\twocolumn[\hsize\textwidth\columnwidth\hsize\csname @twocolumnfalse\endcsname

\title{Optical and Magneto-optical Response of a Doped Mott Insulator}
\author{Mukul S. Laad$^1$, Luis Craco$^2$, and E. M\"uller-Hartmann$^1$}

\address{${}^1$Institut fuer Theoretische Physik, Universitaet zu Koeln, 
Z\"ulpicher Strasse, 50937 Koeln, Germany \\
         ${}^2$Instituto de F\'{\i}sica Gleb Wataghin - UNICAMP, 
C.P. 6165, 13083-970 Campinas - SP, Brazil}
\date{\today}
\maketitle

\widetext

\begin{abstract}
We study the optical, Raman, and ac Hall response of the doped Mott 
insulator within the dynamical mean-field theory ($d=\infty$) for strongly 
correlated electron systems. The occurence of the {\it isosbectic} point 
in the optical conductivity is shown to be associated with the frequency 
dependence of the generalized charge susceptibility.  We compute the Raman 
response, which probes the fluctuations of the "stress tensor", and show 
that the scattering is characterized by appreciable incoherent contributions.
The calculated ac Hall constant and Hall angle also exhibit the isosbectic 
points.  These results are also compared with those obtained for a 
{\it non-FL} metal in $d=\infty$. The role of low-energy coherence (FL) or 
incoherence (non-FL) in determining the finite frequency response of strongly 
correlated metals in $d=\infty$ is discussed in detail.

\end{abstract}
\noindent


]

\narrowtext
\section{INTRODUCTION}

The celebrated Landau theory of the Fermi liquid (FL)~\cite{AGD}, has 
been the mainstay of the conventional theory of metals for over four decades.
It has proved to be remarkably stable even when local correlations are 
strong, as has been observed in a variety of heavy fermion metals.  Recent
theoretical work concentrating on the $d=\infty$ limit of lattice fermionic 
models has shown up the robustness of the local Fermi liquid picture, even 
for large values of the interaction~\cite{GKKR}.

The $d=\infty$ works have also clarified the conditions under which the
metallic phase for a given lattice model is described by FL 
theory~\cite{GK}; in the situation where symmetry breaking is suppressed, 
the FL metal survives as long as there is no ground state degeneracy, as in 
the Hubbard model. In every case where the symmetry-unbroken ground state exhibits a degeneracy 
as a function of the model parameters, FL behavior is invalidated; examples 
are the spinless Falicov-Kimball model~\cite{GKS} and the two-channel Kondo 
model~\cite{NP} in this limit. The crucial role of the transfer of spectral 
weight over wide energy scales across the insulator-metal transition has 
also been revealed; it has thereby become clear that the itinerant and the 
atomic aspects of the problem should be treated consistently on a common 
footing to obtain reliable answers.

Optical conductivity is an insightful probe which can provide detailed 
information about the finite frequency charge dynamics of a correlated 
electronic fluid~\cite{see1}.  In recent years, much information about 
the non-Fermi liquid charge dynamics in the "normal" state of cuprate 
superconductors has been gleaned by careful optical 
measurements~\cite{see1}. The inplane optical conductivity, 
$\sigma_{ab}(\omega)$ for hole doped cuprates shows a non-FL fall-off with 
$\omega$ at small $\omega<<D$ ($D$ is the bandwidth)~\cite{see1}, while
the electron-doped cuprate $Nd_{2-x}Ce_{x}CuO_{4-y}$ shows behavior 
characteristic of a FL~\cite{see1}.  A curious feature observed in these 
studies is the fact that the various $\sigma_{ab}(\omega)$ curves for 
different dopings cross at a {\it single} point (this is the {\it isosbectic} 
point in the literature).  It is interesting to ask for the underlying 
physics manifesting itself in such observations.

Hall measurements provide a detailed picture of finite frequency charge 
excitations in an external magnetic field, and probe the nature of the 
transverse scattering processes in an electronic fluid~\cite{Lange}.  
Measurements carried out on the cuprates show that the transverse scattering 
processes are characterized by anomalous temperature and frequency 
dependences, leading to the two-relaxation rate phenomenology~\cite{Lange}.  
It is well known that FL transport is generically characterized by a 
{\it single} scattering rate governing transport, and the appearance of 
two relaxation rates is therefore cited as a striking manifestation of the 
breakdown of FL ideas.

Given the above, the theoretical problem of computing the ac conductivity
tensor for a strongly correlated metal is interesting; this quantity encodes
detailed information about the finite frequency charge excitations and their 
response to external electric and magnetic fields. Theoretically, the problem 
of computing the transport coefficients for a strongly correlated fermionic 
system in a controlled way is a rather hard task. The problem is even harder 
in the case of magnetotransport; to compute the Hall conductivity tensor, 
one has to evaluate explicitly three-point functions. In finite spatial 
dimensions, vertex corrections, which may be important, cannot be evaluated 
satisfactorily in any controlled approximation.  The above difficulties make 
it imperative to search for controlled approximation schemes where some of 
the above difficulties can be circumvented without sacrificing essential 
correlation effects.       

The dynamical mean-field approximation (DMFA or $d=\infty$) has proved to be
a successful tool to investigate transport in strongly correlated systems in 
a controlled way~\cite{Khurana}.  This is because the vertex corrections 
entering in the Bethe-Salpeter eqn for the two-particle propagator for the 
conductivity vanish rigorously in $d=\infty$.  To evaluate the conductivity 
tensor, one needs only to compute the fully interacting {\it local} 
self-energy of the given model, following which the Kubo formalism can be 
employed~\cite{Moller}.  Given that DMFA captures the nontrivial local 
dynamics exactly, one expects that it provides an adequate physical 
description in situations where local fluctuations are dominant. In fact, 
the dc resistivity and Hall effect, as well as the longitudinal ac 
conductivity for the Hubbard model, have already been considered in the 
literature, in the framework of the $d=\infty$ approximation~\cite{Majumdar}.

However, the harder problem of computing the frequency dependent Hall effect
(Hall constant and angle) has not been studied to date in detail.  Actually,  
Lange~\cite{Lange} has employed the Mori-Zwanzig projection formalism to 
study dc magnetotransport in the Hubbard model.  The actual evaluation of the
complicated equations is, however, actually carried out in the Hubbard I 
approximation. This is known to lead to spurious instabilities (like 
ferromagnetism, which is washed away when local quantum fluctuations 
are included).  Moreover, only the dc Hall constant is evaluated explicitly.  
The Hubbard I approximation does not correctly capture the transfer of 
high-energy spectral weight to low energy upon hole doping, a feature 
characteristic of correlated systems, and so one expects that it will be 
inadequate when one attempts to look at the ac conductivity.  Some of the 
deficiencies of the Hubbard I approximation maybe cured by the so-called 
Hubbard III approximation~\cite{LC}; this is actually the exact solution of 
the $d=\infty$ Falicov-Kimball model~\cite{CG}.  However, the metallic phase 
is not a Fermi liquid, and actually describes the charge dynamics in a model 
with X-ray edge singularities at low energy~\cite{GKS}. More recently, 
Lange {\it et al.}~\cite{Lange} have computed the ac Hall constant using the 
IPT in $d=\infty$.  However, the ac Hall angle and the Raman response, which 
we compute, has not been considered in Ref.~\cite{Lange}.  Additionally, we
have also compared our results with those obtained for a non-FL 
metal~\cite{LC} in $d=\infty$ to clarify the role of low energy coherence 
(incoherence) in determining the ac response of correlated metals.  
The detailed doping and frequency dependence of the Hall constant 
and Hall angle for a strongly correlated FL in $d=\infty$ thus remains an 
open problem to date.

In this paper, we attempt to fill in this gap by studying the ac Hall effect 
and its doping dependence in detail.  We concentrate on the one-band Hubbard 
model; the simplest model exhibiting FL behavior in $d=\infty$.  We have   
recently studied the ac Hall response for the ``simplified Hubbard model''
~\cite{LC} exactly in this limit- as mentioned above, it describes the 
response of a non-FL metal.  We make an in-depth study of all the properties 
which can be computed without making further approximations.  Specifically, 
we study the optical conductivity tensor, the ac Hall constant and Hall
angle, as well as the Raman intensity lineshape as a function of filling for 
the correlated FL metal.  We also compare our results with those obtained 
for a non-FL metal in $d=\infty$.

\section{MODEL AND COMPUTATIONAL DETAILS}
 
We start with the one-band Hubbard model,
\be
H=-t\sum_{<ij>,\sigma}(C_{i\sigma}^{\dag}C_{j\sigma}+h.c) + 
U\sum_{i}n_{i\uparrow}n_{i\downarrow}
\ee
defined on a hypercubic lattice in $d$ dimensions. In the $d\rightarrow\infty$
limit, this corresponds to a gaussian unperturbed DOS~\cite{GKKR}. In this 
limit, the lattice model, eqn.(1), is mapped onto a one-channel Anderson 
impurity problem (SIAM) embedded selfconsistently in a dynamical ``bath'' 
function that encodes the dynamical information about the quantum nature of 
the problem. Solution of the HM (eqn.(1)) therefore requires a reliable way 
to solve the impurity problem~\cite{GKKR}. Unfortunately, there is no exact 
analytical solution available for the SIAM, and one has to resort to schemes 
which give reliable answers, and agree well with "exact" methods, e.g, with 
exact diagonalization~\cite{GKKR}.  In this paper, we use the iterated 
perturbation theory (IPT) away from half-filling~\cite{KK} to solve the SIAM.  
This technique uses the Friedel sum rule~\cite{Lang} (equivalent to the 
Luttinger theorem) to ensure that the correct Fermi liquid behavior is 
recovered at low energy.  By construction, it is also exact in the band and 
the atomic limits, and so is a reliable interpolation scheme that describes 
the {\it full} local dynamical spectrum for all $U/t$ and band-fillings.  
Knowledge of the local dynamics in the SIAM enables us to compute the full 
local dynamics of the HM in $d=\infty$.  We do not repeat the features of 
this method here, but refer the interested reader to Ref.~\cite{KK} for 
details. 

Solution of the $d=\infty$ problem using the IPT yields the {\it full} local
self energy $\Sigma(\omega)$ and the Green function 
$G(\omega)=1/N\sum_{\bf k}G({\bf k},\omega)=1/N \sum_{\bf k} 
[\omega-\epsilon_{\bf k}-\Sigma(\omega)]^{-1}$ 
for the symmetry unbroken paramagnetic case, which we consider here.  We 
have computed the local DOS for the HM, and have checked that all features of 
the strongly correlated Fermi liquid metal are reproduced in accordance 
with~\cite{KK}. The quadratic energy dependence of $Im\Sigma(\omega)$ at 
small energy, the collective Fermi liquid peak that shifts to lower energy 
with hole doping, as well as the transfer of spectral weight from the 
upper- to the (central FL resonance + lower Hubbard band) are all reproduced 
well.

\section{OPTICAL CONDUCTIVITY AND RAMAN LINESHAPE}

\begin{figure}[b]
\epsfxsize=3.5in
\epsffile{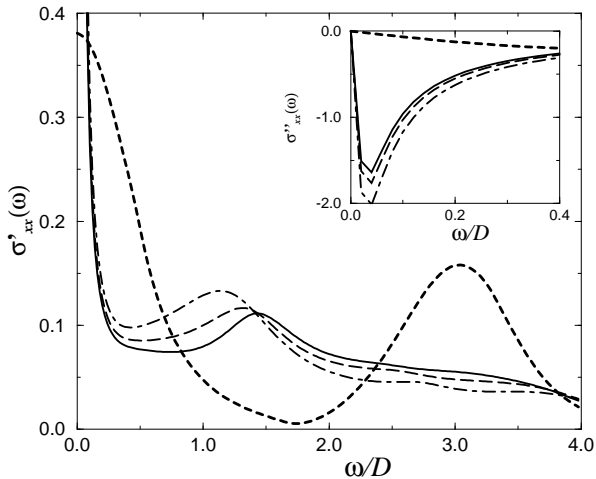}
\caption{The optical conductivity of the Hubbard model with $U/D=3.0$ for 
various band-fillings; $\delta=0.1$ (solid line), $\delta=0.2$ (dashed line),
$\delta=0.3$ (dot-dashed line).  $\sigma_{xx}(\omega)$ for the FKM (non-FL) for
$U/D=3.0$ and $\delta=0.1$ (bold-dashed line).}
\label{fig1}
\end{figure}

As mentioned above, knowledge of $\Sigma(\omega)$ for the lattice problem 
is the only input required to calculate the optical conductivity tensor in 
$d=\infty$.  Since vertex corrections drop out in the two-particle eqn. for the
conductivity in this limit, the longitudinal optical conductivity is given by
a simple bubble diagram involving the {\it full} interacting, local 
propagators. Using the well known eqn. for $\sigma_{xx}(\omega)$~\cite{GKKR}, 

\be
\sigma_{xx}(i\omega)=\frac{1}{i\omega}\int \rho_{0}(\epsilon)\sum_{i\nu}
G(\epsilon, i\nu)G(\epsilon, i\omega+i\nu)
\ee
we have computed $\sigma_{xx}(\omega)$ for a given choice $U/D=3.0$ and for 
different band-fillings $\delta=(1-n)$ ($D$ is the effective bandwidth of the 
non-interacting model). The results are shown in fig.~\ref{fig1}. 
We see that 
$\sigma_{xx}(\omega)$ shows three distinct features worthy of mention: (1) 
The low-energy part ($\omega<0.25$) is characteristic of a renormalized FL 
metal, and correlates well with the behavior of the central FL peak in the 
DOS.  (2) However, $\sigma_{xx}(\omega)$ begins to rise again around 
$\omega/D \simeq 1$, passes through a broad peak centered around $U/2$ 
(for $n=1$), and starts falling off at larger frequencies.
The transfer of optical spectral weight from high- to low-energy upon hole 
doping is clearly exhibited, and is understandable in terms of the interplay 
between the high energy (localized) part, which inhibits double occupancies, 
and the low energy (coherent) part of the spectrum, reflecting carrier 
itinerancy that increases with hole doping.  (3) Most interestingly, the 
$\sigma_{xx}(\omega)$ curves for various $\delta$ all cross at a {\it single} 
point ($\omega \simeq 1.4D$); this is the {\it isosbestic} point cited in 
earlier work~\cite{GKS}.  We have a partial understanding of this curious 
feature: Following Vollhardt~\cite{Vollh}, who investigated such features 
seen in the specific 
heat for the HM for different $U$, we ask for the reasons for the crossing of 
the different curves, and the width of the crossing region.  Focussing on the 
first part of the question, it is easy to show that there must exist some 
$\omega=\omega_{c}(n)$ for which the curves cross.  First, consider the 
high-energy limit of our results.  In general, 
$\sigma_{xx}(\omega)=-\chi"(\omega)/\omega$.  Now, as 
$\omega\rightarrow\infty$, $\Sigma(\omega)=U^{2}(n/2)(1-n/2)/\omega$, and 
$G(\omega)=1/\omega$, so that 
$\sigma_{xx}(\omega\rightarrow\infty)=(n/2)(1-n/2)/\omega$, whereby 
$d\sigma_{xx}/dn > 0$ for $\omega\rightarrow\infty$ for all $n<1$.  On the 
other hand, it is clear from fig.(1), as well as from the optical sum rule, 
that $\sigma_{xx}(\omega)$ 
{\it increases} with increasing $\delta$, i.e with {\it decreasing} $n$, 
at energies $\omega \le 1.4D$, so that we have $d\sigma_{xx}/dn < 0$ at 
small $\omega$.  This implies that $\sigma_{xx}(\omega)$ curves for different 
$\delta$ must cross at some $\omega_{c}(n)$.  This 
crossing occurs at a single point if $\omega_{c}(n)$ is independent of $n$.
Our results show that this seems to be the case to a high accuracy, but we are 
unable to quantify this argument further.  It is an interesting problem to
inquire into the deeper reasons for such features in the HM (see
Ref.~\cite{Vollh} for a thorough discussion of the crossing points in 
$C_{p}(U,T)$).

\begin{figure}[h]
\epsfxsize=3.5in
\epsffile{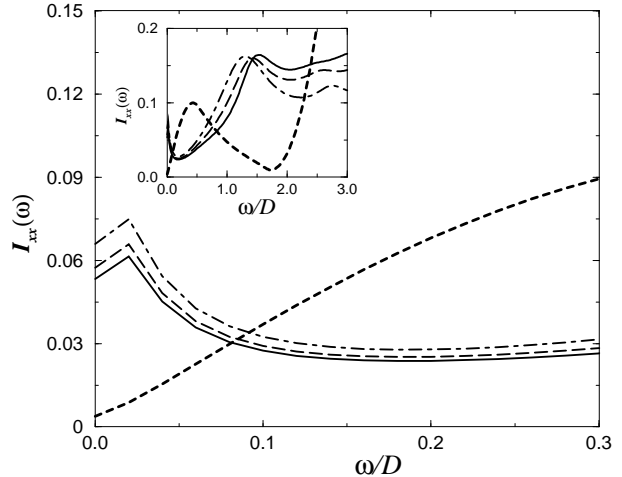}
\caption{Raman intensity lineshape  $I_{xx}(\omega)$ for the Hubbard model 
(FL) with $U/D=3.0$, and for various band-fillings; $\delta=0.1$ (solid line),
$\delta=0.2$ (dashed line), $\delta=0.3$ (dot-dashed line).  $I_{xx}(\omega)$ 
for the FKM (non-FL) for $U/D=3.0$ and $\delta=0.1$ (bold-dashed line).} 
\label{fig2}
\end{figure}

Knowledge of the longitudinal optical conductivity allows us to compute the
Raman spectrum within the $d=\infty$ methodology.  This is because, as long as
$\sigma_{xx}(\omega)$ is completely determined by the renormalized bubble 
contribution (vertex corrections drop out, as in $d=\infty$), the Raman 
intensity is directly related to the optical conductivity~\cite{SS} via the 
eqn.,
\be
I_{xx}(\omega)=\frac{\omega}{1-e^{-\beta\omega}} Re \sigma_{xx}(\omega)
\ee

In fig.~\ref{fig2}, we show the Raman lineshapes calculated from the above 
eqn. for the same $U/D$ and doping values.  Since all calculations are 
carried out at $T=0.01$, $I_{xx}(\omega)$ is finite at $\omega=0$.  A sharp 
peak at low energy, reminiscent of the plasmon peak in continuum 
treatments~\cite{SS}, is visible, followed by an incoherent response 
characteristic of Mott-Hubbard systems~\cite{SS} at higher energies.  
However, the low energy peak is more the characteristic of
the coherent part of the particle-hole response at low energies in a strongly
correlated Fermi liquid.  Evidence in favor of this interpretation is 
provided by the fact that the sharp peak broadens out as the temperature is 
raised above $T_{K}$, the lattice Kondo temperature, where it is destroyed by
strong scattering off local moments (which exist for $T>T_{K}$, rendering the
FL description invalid).

It is interesting to compare the above results with those obtained for a 
non-Fermi liquid (NFL) metal in $d=\infty$.  In this limit, NFL behavior at
$T=0$ is obtained when the local Kondo effect is suppressed; this is exactly
what happens in the Falicov-Kimball model (FKM) in this limit~\cite{GKS}.  
Since the $\downarrow$-spin hopping is zero in the FKM, the Kondo effect 
does not occur, and the metallic state is not a FL near half-filling (it 
is actually described by the local X-ray edge physics in $d=\infty$).  In 
the inset of figs.~\ref{fig1}-\ref{fig2}, we show the real part of 
$\sigma_{xx}(\omega)$ and the Raman intensity lineshape for the FKM with 
identical values of $U/D=3.0$ and hole doping ($\delta=0.1$). It is seen 
that $\sigma_{xx}(\omega)$ for the FKM falls off much more slowly (like 
$1/\omega$) in comparison with the fast Drude-like fall off observed for 
the HM. As expected, the sharp low energy peak in $I_{xx}(\omega)$ is 
changed into a broad continuum, reflecting an incoherent low-energy response 
characteristic of a NFL metal.  We also observe that $I_{xx}(\omega=0)\ne 0$ 
at $T=0$ in the FKM case, as argued by Shastry {\it et al.}~\cite{SS}. This 
corresponds to the fact that, in a
non-FL metal (where the Green fn. has a branch cut rather than a pole 
structure), the action of the kinetic energy, or the stress tensor (which has
non-vanishing matrix elements between lower Hubbard band (LHB) states) does 
not create well-defined elementary excitations.  This results in the broad 
continuumwith non-vanishing intensity in the FKM (NFL), in contrast to the 
sharp peak with vanishing intensity (at $T=0$) in the HM (FL) observed above. 
  Hence, the finite frequency response of correlated metals is determined by
the presence of a pole (branch cut) structure in 
the single-particle Green fn. at
low energy.  Within the $d=\infty$ ideas used here, the low energy coherence is
a manifestation of the collective lattice Kondo effect in the Hubbard model, 
while the incoherent response in the non-FL case is understood in terms of the
vanishing of the $\downarrow$-spin hopping, and to the complete suppression of
this Kondo scale.

\section{AC HALL CONSTANT AND ANGLE}

As mentioned above, computation of the magnetotransport is an extremely 
delicate matter, since one has to evaluate {\it three-point} 
functions~\cite{Lange} to first order in the external magnetic vector 
potential ${\bf A}$. Generally speaking, in a nearly-free electron picture, 
the Hall effect is determined by the vagaries (shape and size) of the Fermi 
surface. That such a correspondence cannot be made for strongly correlated 
metals was pointed out by Shastry {\it et al.}~\cite{SSS} who showed that 
the Hall constant in a strongly correlated system is dominated by
spectral weight far from the Fermi surface, and hence is independent of its
shape.  This suggests that the results should not be sensitive to the choice of
the free DOS, allowing us to use the $d=\infty$ DOS for the hypercubic lattice.
Calculations for a three-dimensional system, where the $d=\infty$ approach 
works quite well~\cite{GKKR}, can be carried out by replacing the $d=\infty$ 
DOS by a 3d DOS.
 
To compute the Hall conductivity, we need to add a Peierls coupling term to 
the HM, where it enters via the hopping~\cite{SSS}. The Hamiltonian in a 
magnetic field is

\be
H=-\sum_{<ij>,\sigma}t_{ij}({\bf A})(C_{i\sigma}^{\dag}C_{j\sigma}+h.c) + 
U\sum_{i}n_{i\uparrow}n_{i\downarrow}
\ee
where the hopping matrix elements are modified by a Peierls phase factor and 
are \mbox{$t_{ij}=exp(2i\pi/\phi_{0}\int_{i}^{j}{\bf A}\cdot d{\bf l})$}, 
where {\bf A} is the vector potential and $\phi_{0}=hc/e$. The off-diagonal 
part of the conductivity involves the computation of a {\it three-point} 
function to first order in the external field, as mentioned before. 
Fortunately, a convenient form has been worked out by Lange~\cite{Lange}, 
so we use the approach developed there.  Explicitly,after a somewhat tedious 
calculation, the imaginary part of $\sigma_{xy}(\omega)$ is given by

\begin{eqnarray}
\nonumber
\sigma_{xy}^{"}(\omega) &= & c_{xy}\int_{-\infty}^{+\infty} 
d\epsilon
\rho_{0}(\epsilon)\epsilon \int_{-\infty}^{+\infty} d\omega_{1}
d\omega_{2} A(\epsilon,\omega_{1})A(\epsilon,\omega_{2})
\\ &&\frac{1}{\omega} \left[
\frac{F(\epsilon,\omega_{1};\omega)-
F(\epsilon,\omega_{2};\omega)}{\omega_{1}-\omega_{2}}
+(\omega \rightarrow -\omega) \right]
\end{eqnarray}
where

\be
F(\epsilon, \omega; \omega_{1}) = A(\epsilon, \omega_{1}-\omega)
[f(\omega_{1})-f(\omega_{1}-\omega)]
\ee
and $A(\epsilon,\omega)
=-Im [\omega-\epsilon-\Sigma(\omega)]^{-1}/\pi$ is the
s.p spectral function in $d=\infty$. 

  Given the s.p spectral fn. in $d=\infty$, $\sigma_{xy}"(\omega)$ is 
computed from the above, and the corresponding real part is obtained from a 
Kramers-Kr\"onig transform.  The Hall constant and Hall angle are obtained 
directly as,
\be
R_{H}(\omega)=\frac{\sigma_{xy}(\omega)}{\sigma_{xx}^{2}(\omega)}
\ee
and,
\be
cot \theta_{H}(\omega)=\frac{\sigma_{xx}(\omega)}{\sigma_{xy}(\omega)}
\ee

\begin{figure}
\epsfxsize=3.5in
\epsffile{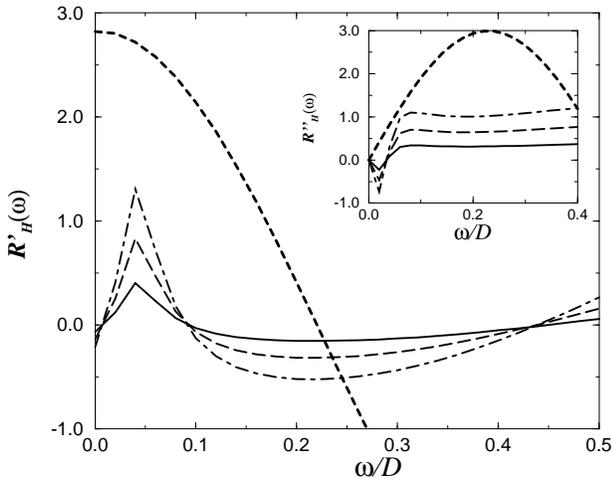}
\caption{AC Hall constant of the Hubbard model with $U/D=3.0$ for 
various band-fillings; $\delta=0.1$ (solid line), $\delta=0.2$ (dashed line),
$\delta=0.3$ (dot-dashed line).  $R_{H}(\omega)$ for the FKM (non-FL) for
$U/D=3.0$ and $\delta=0.1$ (bold-dashed line).}
\label{fig3}
\end{figure}

We describe the results obtained for the HM with $U/D=3.0$ and 
$\delta=0.1,0.2, 0.3$. In fig.~\ref{fig3}, we show the real part of the 
ac Hall constant, $R_{H}'(\omega)$, with the corresponding imaginary part in 
the inset. Since we work at a small $T=0.01$, the $\omega=0$ part of 
$R_{H}'$ is finite and negative, in agreement with~\cite{Majumdar} 
($R_{H}''$ is identically zero at $\omega=0$).  A sharp low energy peak, 
whose strength grows with increasing doping, is clearly seen.  It is 
associated with {\it intraband} transitions within the lower Hubbard band.  
This is related to the fact that the kinetic energy has matrix elements 
between the lower Hubbard band (LHB) states, which correspond to 
quasicoherent processes with weight increasing with $\delta$. Interestingly 
enough, the isosbectic behavior shown in the optical conductivity is also 
shown up in $R_{H}'(\omega)$.  The corresponding results for the FKM (NFL) 
are shown with the bold-dashed line in fig.~\ref{fig3}. Two features are 
worthy of mention. 
(1) $R_{H}'(\omega=0)>0$, in contrast to what is observed for the HM (FL), 
where it is negative, and (2) the low energy peak in the HM is replaced by 
a smoothly decreasing Hall response as expected. The underlying reason for 
this is again that in the FKM, the kinetic energy (stress tensor), which 
connects LHB states, does not create well-defined elementary excitations 
(since the Green fn. is totally incoherent).

\begin{figure}[htb]
\epsfxsize=3.5in
\epsffile{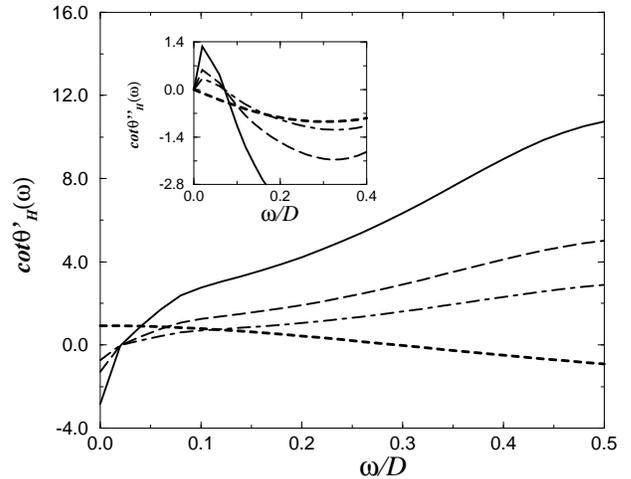}
\caption{AC Hall angle, $cot \theta_{H}(\omega)$ for the Hubbard model with 
$U/D=3.0$ for various band-fillings; $\delta=0.1$ (solid line), $\delta=0.2$ 
(dashed line), $\delta=0.3$ (dot-dashed line).  $cot \theta_{H}(\omega)$ for 
the FKM (non-FL) for $U/D=3.0$ and $\delta=0.1$ (bold-dashed line).}
\label{fig4}
\end{figure}

In fig.~\ref{fig4}, we show the real and imaginary parts of the ac Hall 
angle computed from the above eqn.  It is clear that the strong frequency 
dependence of $cot \theta_{H}(\omega)$ found near half-filling ($\delta=0.1$) 
is weakened on increasing hole doping.  The imaginary part is zero for 
$\omega=0$.   This is again in marked contrast to the results of a similar 
calculation done for the NFL case (FK model), results for which are shown 
with the bold-dashed line in fig.~\ref{fig4}.  We have also computed the 
frequency dependent 
transverse scattering rate, $\Gamma_{xy}(\omega)$ which goes like
$\omega^{2}$ in the FL case.  Since the transport scattering rate as deduced
from the Drude-like response of $\sigma_{xx}(\omega)$ also varies quadratically
in the $d=\infty$ Hubbard model, the longitudinal and transverse scattering 
rates are described by a {\it single} relaxation rate, as they should be in a 
FL metal.  This is in contrast to the results obtained from a similar analysis 
for the FKM~\cite{LC}, where the longitudinal and transverse responses to an 
applied Lorentz field are governed by qualitatively different timescales.

An amusing feature of the present calculations done for the Hubbard model is
the existence of isosbectic points in various calculated quantities, like
$\sigma_{xx}'(\omega)$, $R_{H}(\omega)$, and $cot \theta_{H}(\omega)$, as 
a fn. of hole doping (see figs.(\ref{fig1}-\ref{fig4})). It is 
an extremely interesting problem to examine this feature in more detail, 
perhaps in a manner analogous to Ref.~\cite{Vollh}.
We leave this issue for a more detailed investigation in future.

A detailed investigation of the magneto-optical response for the $d=\infty$
Hubbard model, which in principle is computable from the conductivity tensor,
will be reported separately.

\section{CONCLUSIONS}

In conclusion, we have determined the optical, Raman, and ac Hall response of
the $d=\infty$ Hubbard model using the IPT off half-filling as a reliable 
approximation~\cite{KK}.  We have identified interesting features in the 
results (isosbectic points mentioned above), and shown that the response of 
a strongly correlated Fermi liquid metal is explainable in terms of the 
competition between the atomic and itinerant aspects inherent in the Hubbard 
model. The evolution of the ac response with hole doping is controlled by 
the increasing weight of the quasicoherent processes (corresponding to 
transitions involving only the lower Hubbard band states) relative to that 
of the high-energy incoherent features, whose weight diminishes with 
progressive hole doping. Lastly, we have compared our results with those 
computed earlier by us for a non-FL metal in $d=\infty$, and have discussed 
the role of low-energy coherence (or incoherence) in determining finite 
frequency response of strongly correlated metals. Given the 
success~\cite{GKKR} of the $d=\infty$ methodology in understanding aspects 
of the physics of three-dimensional transition-metal oxides, we believe 
that the calculation presented here in combination with the actual 3d 
bandstructure (which is equivalent to using the LDA DOS, for e.g) 
can be more relevant for studying ac response of 3d TM oxides.

\section{ACKNOWLEDGEMENTS}

One of us (MSL)   
acknowledges financial support of SFB 341.
LC was supported by the Funda\c c\~ao de Amparo 
\`a Pesquisa do Estado de S\~ao Paulo (FAPESP).


\begin{references}

\bibitem{AGD} A. A. Abrikosov, L. P. Gorkov, and I. E. Dzyaloshinskii, in
{\it Methods of Quantum Field Theory in Statistical Physics}, Pergamon, 
Elmsford N. Y.  

\bibitem{GKKR} A. Georges {\it et al.}, Rev. Mod. Phys. {\bf 68}, 
13 (1996).

\bibitem{GK} A. Georges and G. Kotliar, Phys. Rev. B {\bf 45}, 6479 (1992).

\bibitem{GKS} A. Georges, G. Kotliar and Q. Si, 
{\it Int. J. Mod. Phys.} B {\bf 6}, 705 (1992).

\bibitem{NP} P. Nozieres and D. Pines, Phys. Rev. {\bf 109}, 741 (1958).

\bibitem{see1} see for e.g, the Volumes, {\it High Temperature 
Superconductivity}, edited by D. M. Ginsberg.  

\bibitem{Lange} E. Lange, Phys. Rev. B {\bf 55}, 3907 (1997).

\bibitem{Khurana} A. Khurana, Phys. Rev. Lett {\bf 64}, 1990 (1990).

\bibitem{Moller} G. M\"oller, A. E. Ruckenstein, and S. Schmitt-Rink, 
Phys. Rev. B {\bf 46}, 7427 (1992).

\bibitem{Majumdar} P. Majumdar and H. R. Krishnamurthy, preprint 
cond-mat/9512151.

\bibitem{LC} M. S. Laad and L. Craco, preprint cond-mat/9806076, submitted 
to Phys. Rev. B.

\bibitem{CG} U. Brandt and C. Mielsch, Z. Phys. {\bf 75}, 365 (1989);
for different approaches, see also M. S. Laad, Phys. Rev. B {\bf 49}, 
2327 (1994); L. Craco and M. A. Gusm\~ao,  Phys. Rev. B {\bf 54}, 1629
(1996). 

\bibitem{KK} H. Kajueter and G. Kotliar, Phys. Rev. Lett. {\bf 77}, 131 
(1996). 

\bibitem{Lang} D. C. Langreth, Phys. Rev. {\bf 150}, 516 (1960).

\bibitem{Vollh} D. Vollhardt, Phys. Rev. Lett. 78, 1307 (1997).

\bibitem{SS} B. S. Shastry and B. Shraiman, Phys. Rev. Lett. {\bf 65}, 
1068 (1990).

\bibitem{SSS} B. S. Shastry, B. I. Shraiman, and R. R. P. Singh, Phys. Rev. 
Lett. {\bf 70}, 2004 (1993).

\end{references}
\end{document}